# Spectroscopic ellipsometry of graphene and an exciton-shifted van Hove peak in absorption.

V. G. Kravets, A. N. Grigorenko[*], R. R. Nair, P. Blake, S. Anisimova,

K. S. Novoselov, and A. K. Geim.

*School of Physics and Astronomy, University of Manchester, Manchester, M13 9PL, UK.*

**Abstract:** We demonstrate that optical transparency of any two-dimensional system with a symmetric electronic spectrum is governed by the fine structure constant and suggest a simple formula that relates a quasi-particle spectrum to an optical absorption of such a system. These results are applied to graphene deposited on a surface of oxidized silicon for which we measure ellipsometric spectra, extract optical constants of a graphene layer and reconstruct the electronic dispersion relation near the *K* point using optical transmission spectra. We also present spectroscopic ellipsometry analysis of graphene placed on amorphous quartz substrates and report a pronounced peak in ultraviolet absorption at 4.6 eV because of a van Hove singularity in graphene's density of states. The peak is downshifted by 0.5 eV probably due to excitonic effects.

**I. Introduction.**

Graphene, a recently isolated two-dimensional carbon-based material [1], attracted a lot of attention due to unique physical, chemical and mechanical properties [2]. Its low energy excitations, known as massless Dirac fermions [2], result in distinctive properties such as the integer and fractional quantum Hall effect [3-6], minimum metallic conductivity, Klein tunneling, etc. [2]. Recently, it was shown that graphene also possesses remarkable optical properties: its visible transparency is determined by the fine structure constant [7]; photoresponse can reach THz frequencies [8]; infrared transmission can be modulated by a gate voltage [9]. This opens prospects of graphene applications in optics and optoelectronics, e.g., in solar cells [10], liquid crystal displays [11], filters and modulators.

Despite a rapid progress in studying infrared properties of graphene layers [9, 12], works on optical properties of graphene in visible are scarce [13]. The aim of this work is to elucidate optical properties of graphene layers in visible light using variable angle spectroscopic ellipsometry and large (typical size of 100μm) high quality flakes deposited on an oxidized silicon substrate and a transparent quartz substrate. Graphene's absorption remains universal up to violet-light frequencies with a value of 2.3% [7] but exhibits a pronounced peak reaching ~10% in ultraviolet. We also describe how a low energy electronic spectrum of any 2D system with symmetric electronic spectrum can be reconstructed from its optical transmission.

**II. Theory.**

We start by considering optical absorption of a generic 2D system described by Hamiltonian $\hat{H}_0 = \sum_{\vec{p}} \varepsilon(\vec{p}) a_{\vec{p}}^{\dagger} a_{\vec{p}}$. We suppose that the system has a symmetric electronic energy spectrum of quasiparticles such that both $\varepsilon(\vec{p})$ and $-\varepsilon(\vec{p})$ are eigenvalues. In electromagnetic field $\vec{p}$ is gauged as $\vec{p} \to \vec{p} - e\vec{A}/c$, which gives a semiclassical interaction Hamiltonian as $\hat{H}_{int} = -\dfrac{e}{2c}\dfrac{\partial \hat{H}_0}{\partial \vec{p}}\left(\vec{A}\exp(\imath\omega t) + \vec{A}\exp(-\imath\omega t)\right)$, where $\vec{A}$ is the real vector potential of electromagnetic wave of frequency $\omega$ with electric field $\vec{E}$ ($E_0$ being a real amplitude of the electric field in the 2D layer). The energy absorption of normal incident light per unit area is given by the Fermi golden rule:

$$W = \hbar\omega \frac{2\pi g}{\hbar} \sum_{i,f} \left|\langle f|\hat{V}|i\rangle\right|^2 \left(n(\varepsilon_i) - n(\varepsilon_f)\right) \delta(\vec{p}_f - \vec{p}_i)\delta(\varepsilon_f - \varepsilon_i - \hbar\omega), \qquad (1)$$

where $g$ is the degeneracy of states, $i, f$ are the initial and final states, respectively, $\hat{V} = -\dfrac{e}{2c}\dfrac{\partial \hat{H}_0}{\partial \vec{p}}\vec{A}$, and $n(\varepsilon)$ is the occupancy number. For direct optical transitions in a system with a symmetric spectrum (both $\varepsilon(\vec{p})$ and $-\varepsilon(\vec{p})$ are eigenvalues) we have $\varepsilon_f = -\varepsilon_i = \hbar\omega/2$. This gives

$$W = \frac{ge^2}{8\pi\hbar^2\omega} \int \left|\langle f|\vec{E}\frac{\partial\varepsilon}{\partial\vec{p}}|i\rangle\right|^2 \left(n(-\varepsilon) - n(\varepsilon)\right)\delta(\varepsilon - \hbar\omega/2) p\, dp\, d\varphi, \qquad (2)$$

where $\varphi$ is the polar angle in the plane. Assuming that the energy spectrum is also isotropic $\varepsilon(\vec{p}) = \varepsilon(p)$, we write $\dfrac{\partial\varepsilon(\vec{p})}{\partial\vec{p}} = \dfrac{d\varepsilon}{dp}\hat{p}$ (with $\hat{p} = \vec{p}/p$ and $p = |\vec{p}|$) and get

$$W = \frac{ge^2 E_0^2}{8\hbar} S\left(n(-\varepsilon) - n(\varepsilon)\right) \frac{d \ln(\varepsilon)}{d \ln(p)}\bigg|_{\varepsilon = \hbar\omega/2}, \qquad (3)$$

where $S = \frac{1}{2\pi} \int |\langle f | \hat{e}\hat{p} | i \rangle|^2 d\varphi$ is the form-factor ($\hat{e}$ is the unit polarization vector). We consider a sufficiently thin layer in which the electric field is constant along the thickness of the layer and the electric field amplitude of the reflected light is small (such condition can be satisfied since the reflection from a layer tends to zero when the thickness of the layer goes to zero). In this case the incident electric field amplitude is close to that in the 2D layer ($E_0$) and the electromagnetic energy that falls onto the system per unit area is approximately $W_0 = \frac{c}{8\pi} E_0^2$. This suggest that for linear polarization ($S=1/2$) the absorption ratio is

$$Abs(\omega) = W / W_0 = \frac{g\pi\alpha}{2} \cdot \left(n(-\varepsilon) - n(\varepsilon)\right) \frac{d \ln(\varepsilon)}{d \ln(p)}\bigg|_{\varepsilon = \hbar\omega/2}. \qquad (4)$$

This expression connects the experimentally measured spectrum of light absorption $Abs(\omega)$ with the electronic dispersion $\varepsilon(p)$. It is worth noting that $\left(n(-\varepsilon) - n(\varepsilon)\right) \frac{d \ln(\varepsilon)}{d \ln(p)} = \left(n(-\varepsilon) - n(\varepsilon)\right) \frac{p}{\varepsilon} \frac{d\varepsilon}{dp}$ is a number, meaning that absorption of any 2D material is governed by the fine structure constant. For fermionic spectrum simple calculations yield:

$$n(-\varepsilon) - n(\varepsilon) = \sinh(\varepsilon/T) / \left(\cosh(\mu/T) + \cosh(\varepsilon/T)\right), \qquad (5)$$

where $\mu$ is the chemical potential and $T$ is measured in the energy units. It is easy to check that $n(-\varepsilon) - n(\varepsilon) \approx 1$ for optical frequencies ($\varepsilon \gg T$) and small doping ($\mu \ll \varepsilon$). For the linear dispersion relation in graphene we have $\frac{d \ln(\varepsilon)}{d \ln(p)} = 1$, $g=2$ (spin degeneracy) and hence optical absorption is $Abs(\omega) \approx \pi\alpha$ in accordance with [7,14] (the double degeneracy of K points in graphene is accounted for by the larger size of the elementary cell that contains 2 carbon atoms). It is interesting to note that the derivative $\frac{d \ln(\varepsilon)}{d \ln(p)} = n$ for a dispersion given by a homogeneous function $\varepsilon \propto p^n$ and hence the optical absorption of such a system is given by $Abs(\omega) \approx n\pi\alpha$. Therefore, the conclusion of Ref. 7 that the fine constant structure defines transparency of graphene can be generalized to any sufficiently thin electronic system.

Formula (4) can be easily inverted to give the electronic spectrum in terms of the light absorption spectrum as

$$p = C \cdot \exp\left( \pi\alpha \int \frac{(n(-\varepsilon) - n(\varepsilon))d\varepsilon}{\varepsilon Abs(2\varepsilon/\hbar)} \right), \qquad (6)$$

where $C$ is a constant and $n(-\varepsilon) - n(\varepsilon)$ is given by (5). For undoped graphene (6) simplifies to

$$p \approx C \cdot \exp\left( \pi\alpha \int \frac{d\varepsilon}{\varepsilon Abs(2\varepsilon/\hbar)} \right). \qquad (7)$$

Expression (6) provides an electronic dispersion of a symmetrical and sufficiently thin system in terms of its absorption spectra measured at the normal angle of incidence (with a replacement $n(-\varepsilon) - n(\varepsilon) = \sinh(\varepsilon/T)/(\cosh(\mu/T) - \cosh(\varepsilon/T))$ eq. (6) can be used for 2D systems with bosonic spectrum.)

**III. Reconstructed electronic spectra for graphene and graphite.**

We now apply (7) to the absorption spectra of graphite and graphene flakes. In order to find their absorption we make use of ellipsometric spectroscopy described in Section IV below (Figs. 3 and 4). Figure 1 shows the reconstructed electronic dispersion relations of graphite and graphene calculated with eq. (7). Both electronic spectra demonstrate a linear dependence at sufficiently low energies (below ~1eV), which supports observation of a signature of massless Dirac fermions with linear dispersion in graphite using angle resolved photoemission spectroscopy [15]. The linear spectrum persists to higher energies in graphene rather than in graphite. The deviation from the linear spectrum is due to non-isotropic corrections to electronics dispersion [16]. Such deviations cannot be taken into account by essentially isotropic inversion expressions (6), (7). Using (2) we can find the energy $\varepsilon_d$ at which the slope of $\varepsilon(p)$ deviates from linear dependence and get an estimate for hopping energy $t \approx 2.9$ eV. This value is close to the literature value of 2.8eV [16].

**IV. Variable angle spectroscopic ellipsometry of graphene and graphite on a silicon substrate.**

To find optical absorption we used spectroscopic ellipsometry measurements. A schematic representation of experimental set-up is presented on Fig. 2(a). Graphene flakes have been prepared on a silicon wafer covered with ≈300nm layer of silicon oxide in order to improve graphene visibility [1,13,17]. The ellipsometric parameters $\Psi$ and $\Delta$ (defined as $\tan(\Psi)\exp(i\Delta) = r_p / r_s$, where $r_p$ and $r_s$ are the reflection coefficients for the light of *p*- and *s*-polarizations [18]) have been measured with the help of a focused beam variable angle Woollam ellipsometer with a focal spot of just 30μm. Measurements have been performed on the substrate and on relatively large samples, see Fig. 2(b), and modeled with Wvase32 software based on Fresnel coefficients for multilayered films. It is worth noting that the effective optical thickness of thin layers can deviate from "geometrical" thickness of the sample [19].

Indeed, we found that the optical thickness of graphene which gave the best fit changed from flake to flake (varying by about 30% around the value of the interlayer distance of graphite). However, it is impossible to exclude the possibility that the optical thickness of a graphene flake was influenced be residuals on a graphene surface. For this reason, we fixed the thickness of graphene to 0.335nm in our calculations.

We tested our installation on thick flakes of highly ordered pyrolytic graphite. Figure 3(a) and (b) shows ellipsometric spectra of $\Psi$ and $\Delta$ measured on freshly cleaved ~1μm thick graphite for angles of incidence from 45° to 70°. These measured spectra have been fit simultaneously by a Fresnel model in which graphite was described by an anisotropic (uniaxial) material with adjustable parameters $n_x$, $k_x$, $n_z$, $k_z$, where dielectric permittivity $\varepsilon$ is given by $\varepsilon = (n + ik)^2$. (Here we refer to the in-plane optical constants of graphite and graphene as $x$-constants, while the perpendicular to the graphene layer constants - as $z$-constants.). The fit resulted in the reconstructed optical constants shown in Fig. 3(c), the fit itself is shown in Fig. 3(a) and (b) by the white dashed lines.

We note two features of the extracted optical constants. First, there are two regions where spikes appear in the reconstructed spectral dependence of the constants, see the solid arrows at Fig. 3(c). These spikes were induced by instability of the Xe lamp used in the ellipsometer. Second, we found that there is a coupling between reconstructed $x$- and $z$-components (e.g., a strong correlation between $k_x$ and $n_z$, see Fig. 3(c)) which is most likely caused by small variations in flatness of graphite surface and hence deviations of in-plane currents. This coupling might be even stronger for graphene due to intrinsic or extrinsic ripples [20,21] present in graphene sheets. To avoid the coupling we choose to model *z*-response of graphite and graphene flakes as a Cauchy material [18], which is a good approximation for mostly dielectric response of *s*-electrons [22]. Figure 3(d) shows the extracted constants of graphite in this case and Figure 3(a) demonstrates an excellent fit between the calculated (solid blue lines) and the experimental ellipsometric spectra.

As an independent check of our procedure, we calculated the transmission spectrum for a free-standing 0.335nm thick graphite film (corresponding to a graphene monolayer) with the extracted optical constants. It was shown by Kuzmenko *et al.* [14] that the optical transmission for a graphite layer of such thickness should be close to $1-\pi\alpha$, where $\alpha$ is the fine structure constants, at least at the infrared frequencies, see also Ref. 7. Figures 3(e) and (f) show the optical transmission spectra of a 0.335nm thick graphite film calculated with the extracted constants for the set of constants shown in Fig. 3(c) and (d) respectively. We see that indeed the transmission approaches the value $1-\pi\alpha$ at red and near-infrared wavelengths.

The ellipsometry of graphene flakes is shown in Fig. 4. Figure 4(a) demonstrates spectral dependence of $\Psi$ measured at the incidence angle of 45° for the substrate (a green curve), a graphene monolayer (dark yellow), bilayer graphene (maroon), and triple layer graphene (brown). We note an excellent contrast in $\Psi$ at the position of the reflection peaks (~320nm and ~510nm) as a function of the number of layers. The ratio of the maximal $\Psi$ change due to the presence of one graphene layer (~3°) to the $\Psi$ noise (~0.02°) is large. As a result, ellipsometric measurements could provide an efficient way for monitoring the growth of graphene layers on an arbitrary substrate in-situ (with a better accuracy than simple reflection measurements [23]).

The angle dependence of the ellipsometric spectra of a graphene monolayer is shown in Fig. 4(b) and (c). We fit all spectra simultaneously using the multilayered model consisting of silicon substrate, silicon oxide, a Cauchy sublayer and graphene. The thickness of the silicon oxide layer was determined by spectroscopic measurements performed on the bare substrate. A thin Cauchy sublayer under graphene represents the spacer between graphene and the substrate, which usually contains some amount of water and air. (Typical parameters of this layer are: thickness 2-4nm, Cauchy constants A≈1.03, B≈0.08, C≈-4·10$^{-4}$). Graphene was modeled by an anisotropic material of a predefined thickness of 0.335nm with an arbitrary *x*-response and a Cauchy response for *z*-component. The model therefore has only ~2 parameters for a given wavelength which were determined from more than 8 experimental data points (measured at different angles of incidence). The model resulted in an excellent fit to the ellipsometric spectra shown by the white dashed lines on Figs. 4(b) and (c). From this fit we extracted optical constants $n_x$ and $k_x$ plotted on Fig. 4(d) (*z*-component was close to that of graphite). These optical constants describe optical properties of a graphene layer immobilized on an oxidized silicon substrate. We excluded from the data spectral regions near ~380nm and ~800nm where the spikes of the lamp intensity made the extraction procedure unreliable. It is worth nothing that there is a small irregularity in the optical constants of graphene at the wavelength of 540nm (2.3eV), which could be attributed to graphene contamination. The extracted constants have been used for calculation of optical transmission and reconstruction of the free-standing graphene and graphite electronic spectra near the *K* point shown in Fig. 1.

**V. Variable angle spectroscopic ellipsometry of graphene and graphite on an amorphous quartz substrate.**

An accuracy of the extraction of optical constants of graphene could be substantially increased if large high-quality graphene flakes can be made and located on a flat transparent insulating substrate. Indeed, the phase ellipsometric response of a dielectric substrate is trivial ($\Delta=180°$ degree for an angle of incidence smaller than the Brewster angle and $\Delta=0°$ otherwise) and thus any ellipsometric phase variation should be produced by a graphene layer. In addition, the corresponding Fresnel model is simpler since the substrate consists of just one layer whereas a silicon substrate described above also contained 300nm of silicon oxide to insure better graphene visibility.

It is difficult to produce and find sufficiently high quality graphene flakes on a dielectric flat substrate that would allow us to measure the ellipsometric and transmission spectra in wide spectral range directly. Recently we have obtained and located graphene flakes on a dielectric amorphous quartz substrate. The size of the flakes was well in excess of 200x200 $\mu m^2$ which is enough to be measured by a focused beam variable angle Woollam spectroscopic ellipsometer. The quality of the substrate surface allowed us to restore the optical constants in the wavelength range of 240-750nm. At the same time, the reliability of the restoration procedure using Fresnel theory and the accuracy of the calculated optical constants of graphene was much better for an amorphous quartz substrate than those for a silicon substrate.

Figure 5(a) and (b) shows the measured ellipsometric spectra of Ψ and Δ, respectively, from a graphene flake on the amorphous quartz substrate. Figure 5(c) plots the restored optical constants of the graphene layer using Fresnel theory (the amorphous substrate was modeled as a Cauchy materials with parameters A=1.417, B=0.003, C=0 and the graphene layer was modelled as a uniaxial anisotropic material with the thickness of 0.335nm). It is worth noting that Figure 5 provides the first spectroscopic ellipsometry analysis of graphene flakes deposited on a dielectric substrate. Figure 5(d) shows absorption spectra of single-layer graphene as a function of energy $E$ (the solid curves). At visible frequencies, graphene absorbs 2.3% of light but the absorption rapidly grows in violet exhibiting a pronounced peak at 4.6 eV. The peak is due to a van Hove singularity in the density of states, which occurs close to the hoping energy $t$ (see Ref. 16). The dashed curves 1 and 2 in this figure are theory [24] for the non-interaction case and with excitonic effects included, respectively. Two experimental curves, 3 and 4, are shown but all the measured dependences *Abs(E)* fall within the range between these two. Most (~75%) of our samples very closely followed the lower curve that shows minor deviations from 2.3% absorption at visible frequencies, in agreement with Ref. 7. The sample dependence of the onset of the ultraviolet peak remains unclear and could be due to contamination or rippling. At the same time, dozens of the measured samples exhibited the absorption peak precisely at the same energy and as predicted by the case if only the density of states is taken into account (the dashed curve 2).

**VI. Conclusion.**

To conclude, we performed spectroscopic ellipsometry study of graphite and graphene flakes and proposed an inversion formula which calculates the electronic dispersion of a thin symmetric 2D system in terms of an optical absorption spectrum. We applied this formula to both graphite and graphene flakes and demonstrated the existence of linear spectrum of quasiparticles in both cases. We also provided the first spectroscopic ellipsometry measurements of graphene flakes on dielectric substrates and extracted optical constants of graphene layers.

**References.**

Figure captions.

Fig. 1. (Color online) Electronic dispersion relations in graphite and graphene calculated from the optical transmission spectra.

Fig. 2. (Color online) Samples for spectroscopic ellipsometry measurements. (a) Sample geometry. (b) A photograph of a graphene flake.

Fig. 3. (Color online) Variable angle spectroscopic ellipsometry of graphite. (a) and (b) $\Psi$ and $\Delta$ spectra for different angles of incidence $\theta$ measured with a step of 5°. (c) and (d) reconstructed optical constants of graphite with $z$-component being treated as a general material and a Cauchy material, respectively (e) and (f) calculated transmission spectra for 0.335nm thick graphite with constants from (c) and (d) respectively.

Fig. 4. (Color online) Variable angle spectroscopic ellipsometry of graphene on silicon substrate. (a) $\Psi$ spectra measured at $\theta=45°$ for the bare substrate, the substrate with a graphene monolayer, a graphene bilayer and a graphene triple layer. (b) and (c) $\Psi$ and $\Delta$ spectra for different angles of incidence $\theta$ measured on a graphene flake. (d) Reconstructed optical constants of graphene.

Fig. 5. (Color online) Variable angle spectroscopic ellipsometry of graphene on amorphous quartz substrate. [(a) and (b)] $\Psi$ and $\Delta$ spectra for different angles of incidence $\theta$ measured on a graphene flake. (c) Reconstructed optical constants of graphene. (d) Measured absorption spectra of graphene (the solid curves) exhibit a pronounce asymmetric peak in ultraviolet which is shifted from its expected position (the dashed curve 1) by 0.5 eV due to excitonic effects [24].

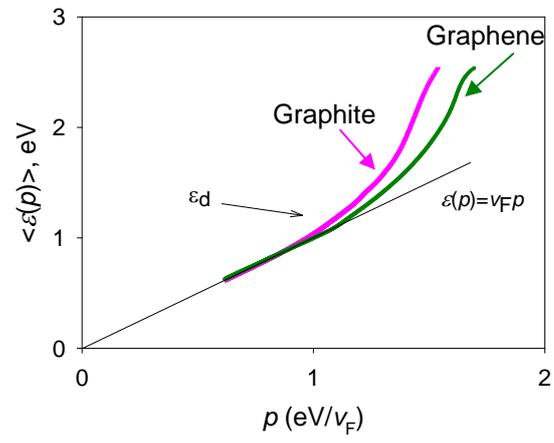

Fig. 1

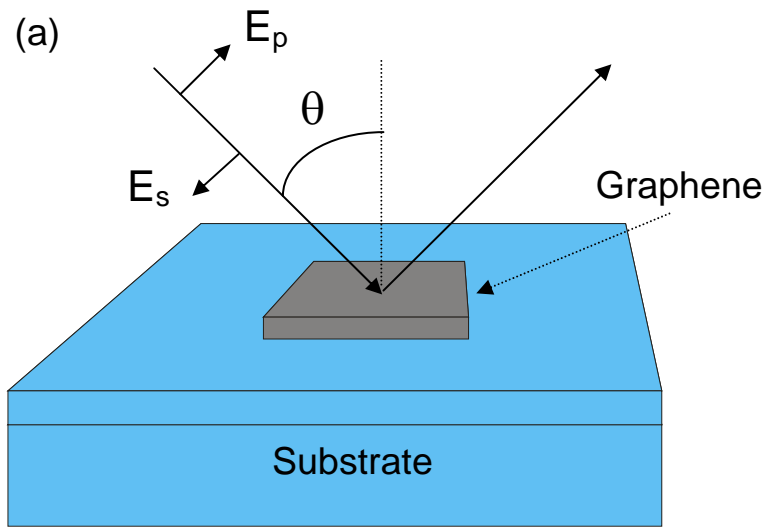

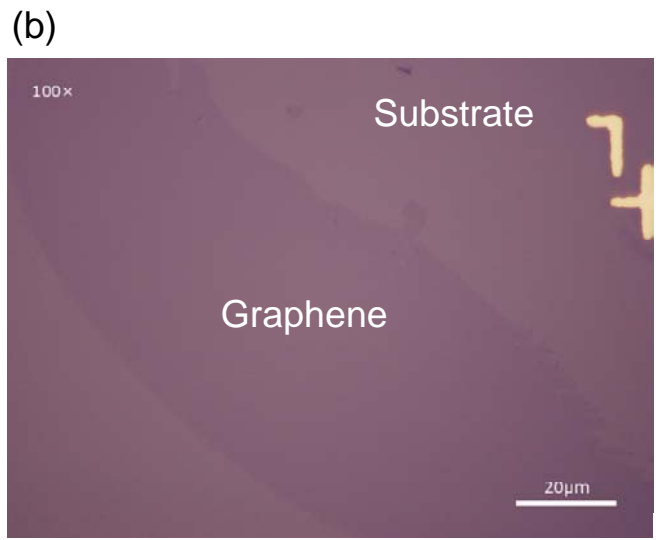

Fig. 2

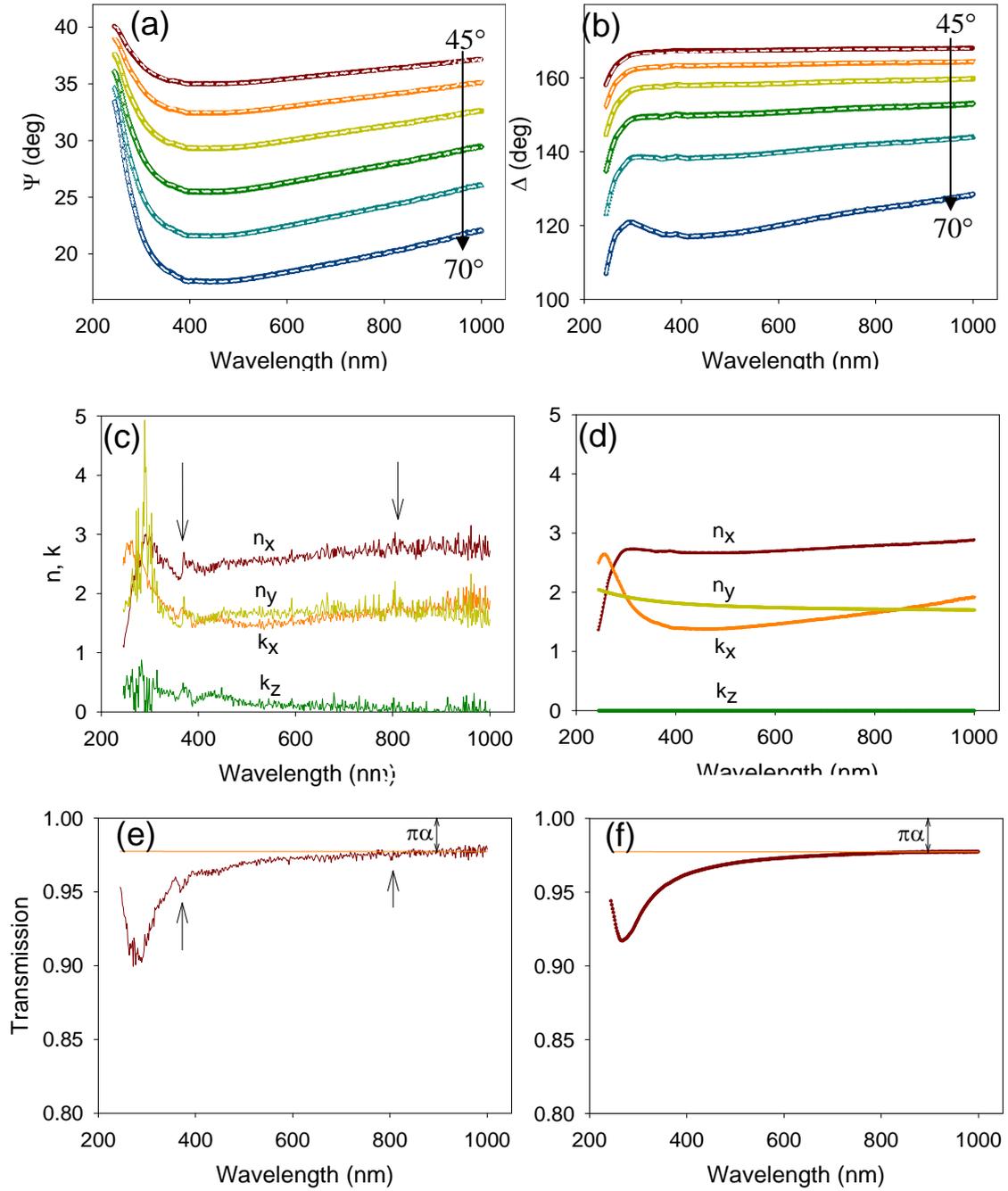

Fig. 3

20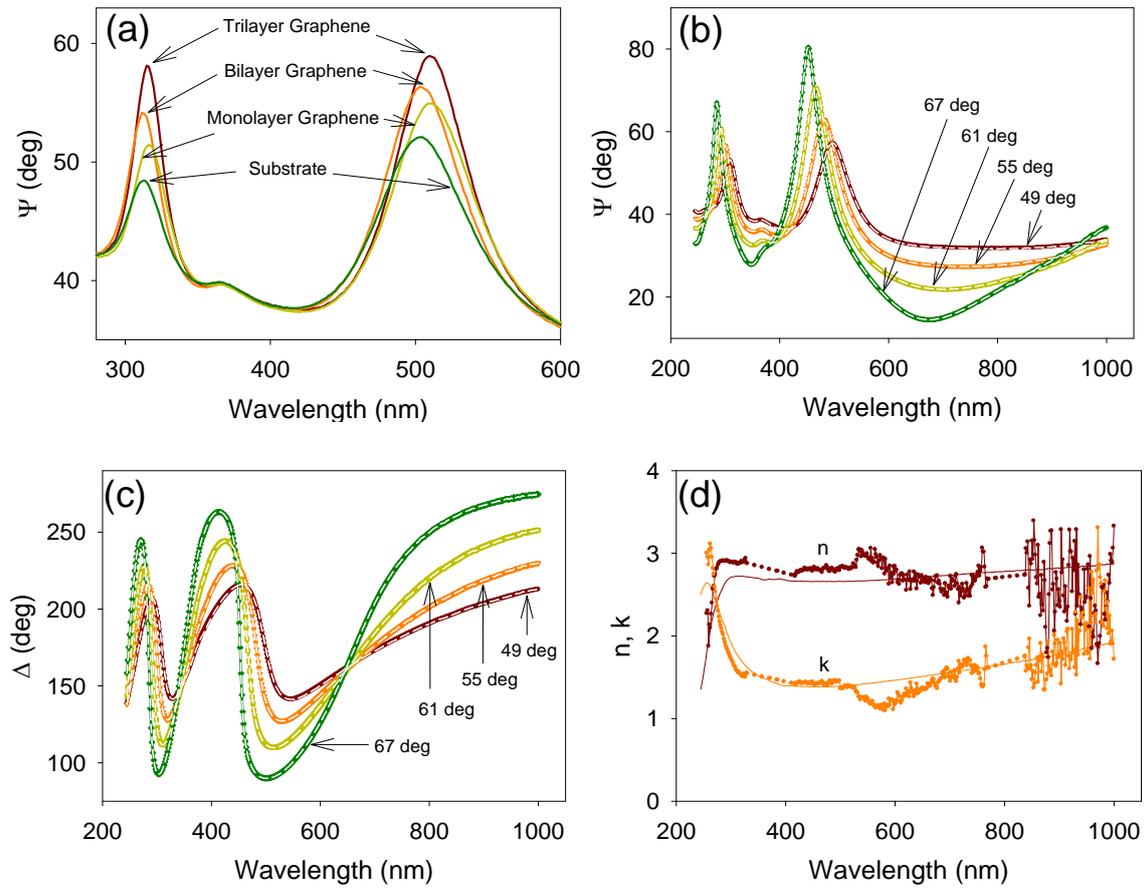

Fig. 4

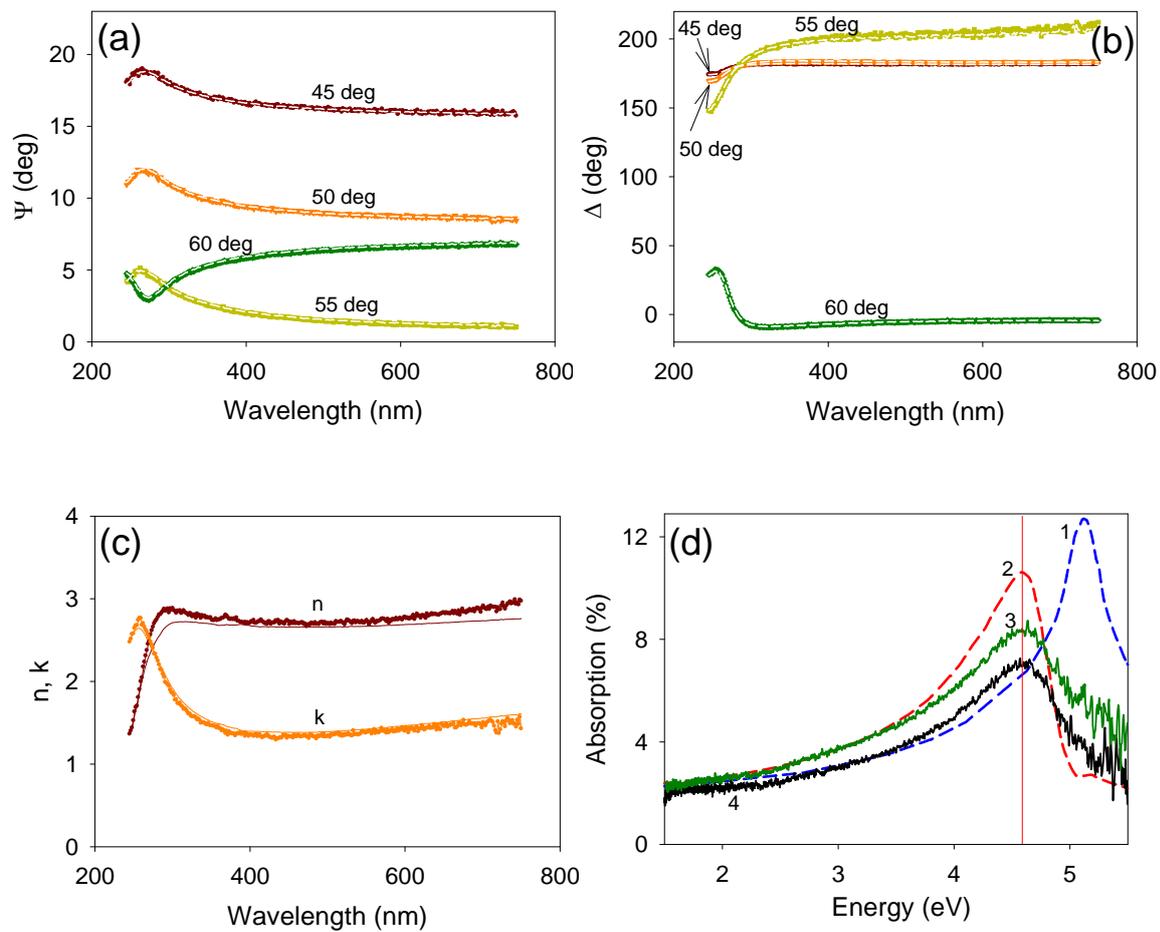

Fig. 5